\documentclass{ws-procs9x6_New}

\begin{document}

\title{Numerical Models of Spin-Orbital Coupling in Neutron Star Binaries}

\author{P. MARRONETTI and S. L. SHAPIRO\footnote{\uppercase{D}epartment of \uppercase{A}stronomy \& \uppercase{NCSA}, \uppercase{UIUC}.}}

\address{Department of Physics, \\
University of Illinois at Urbana-Champaign \\ 
Urbana, IL 61801, USA\\ 
E-mail: pmarrone@uiuc.edu}

\maketitle

\abstracts{
We present a new numerical scheme for solving the initial value problem for quasiequilibrium binary neutron stars allowing for arbitrary spins. We construct sequences of circular-orbit binaries of varying separation, keeping the rest mass and circulation constant along each sequence. The spin angular frequency of the stars is shown to vary along the sequence, a result that can be derived analytically in the PPN limit. This spin effect, in addition to leaving an imprint on the gravitational waveform emitted during binary inspiral, is measurable in the electromagnetic signal if one of the stars is a pulsar visible from Earth.}

We develop a new formalism for the hydrodynamical part of the initial value problem (IVP) that has, as its most important feature, the capability of providing solutions for binaries with stars with arbitrary spins. The gravitational part of the IVP (i.e.; the solution of the Hamiltonian and momentum constraints for a quasiequilibrium circular orbit) is solved using the Wilson-Mathews conformal ``thin-sandwich" approach \cite{Wilson:1995ty}. This method is typically based on restricting the spatial metric tensor to a conformally flat form and imposing a helical Killing vector to the spacetime to enforce the quasiequilibrium ``circular orbit" condition. The details of the new formalism, as well as the numerical code tests, have been presented in Marronetti and Shapiro \cite{Marronetti:2003gk}. We construct sequences of quasiequilibrium orbits with constant rest mass and constant relativistic circulation along the stellar equator. The stars are assumed to be identical and obey an $n=1$ polytropic equation of state. These sequences, which mimic evolutionary inspirals outside the innermost stable circular orbit (ISCO) driven by (slow) gravitational radiation emission, provide insight into the coupling of spin and orbital angular momenta in general relativistic binaries.

\begin{figure}
\epsfxsize=2.8in
\begin{center}
\leavevmode \epsffile{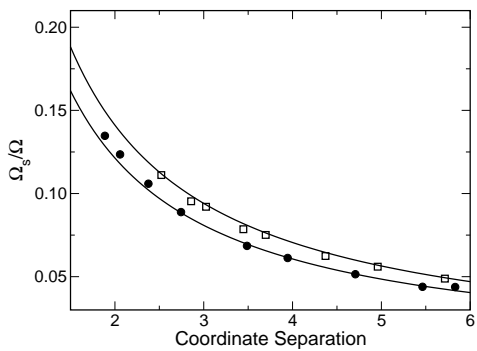}
\end{center}
\caption{ Ratio of spin vs orbital angular velocities as a function of the coordinate separation. The filled circles (open squares) correspond to the zero-circulation sequence with stellar compaction $(m/R)_\infty = 0.14 ~(0.19)$. The solid lines show the analytic PPN prediction (see Appendix F of Marronetti and Shapiro $^2$ for a derivation).}
\label{Spin_vs_d}
\end{figure}

\begin{figure}
\epsfxsize=2.8in
\begin{center}
\leavevmode \epsffile{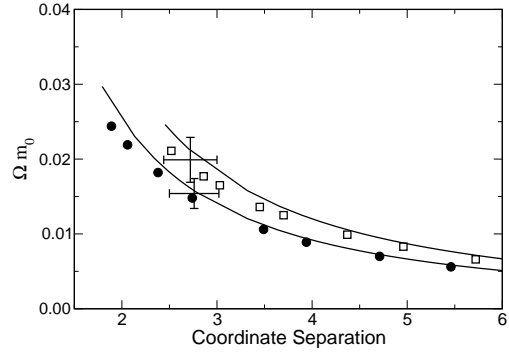}
\end{center}
\caption{ Orbital angular velocities as a function of the coordinate separation for zero-circulation sequences. The caption is identical to Fig. \ref{Spin_vs_d}. The crosses mark the ISCO as determined by the full general relativistic hydrodynamical simulations described in Marronetti {\it et al.} $^3$. The solid lines correspond to a 2.5 PPN expansion, as detailed in Pati and Will $^4$.}
\label{Omega_vs_d}
\end{figure}

\begin{figure}
\epsfxsize=2.8in
\begin{center}
\leavevmode \epsffile{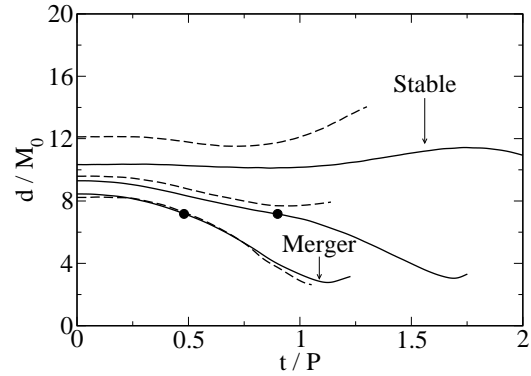}
\end{center}
\caption{ Coordinate separation vs time for binaries with stellar compaction $(m/R)_\infty = 0.14$ and zero-circulation that start at different separations. The curves labeled ``Stable'' (outside the ISCO) and ``Merger'' (inside the ISCO) bracket our estimation of the location of the ISCO. (From Marronetti {\it et al.} $^3$.). Solid (dashed) curves represent high (low) resolutions and the filled circles mark the time of surface contact.}
\label{figure02}
\end{figure}

Figure \ref{Spin_vs_d} shows the evolution of the stellar spin during the inspiral for two zero-circulation (irrotational) sequences. We note an increase of the spin that can reach up to $10\%$ of the orbital angular velocity, before the binary reaches the ISCO. Figure \ref{Omega_vs_d} shows the evolution of the orbital angular velocity. The crosses mark the position of the ISCO as determined by the full general relativistic hydrodynamical simulations described in Marronetti {\it et al.} \cite{Marronetti:2003hx} and shown in Figure \ref{figure02}. We see that the ISCO is located prior to the termination of the quasiequilibrium sequence in both cases. Note also that our results agree very well with post-Newtonian predictions at large separations. This variation of the spin of the star throughout the inspiral is a potentially observable electromagnetic effect when one of the neutron stars is a pulsar. This result takes on added significance in light of the recent discovery \cite{Burgay:2003jj} of the highly relativistic double-pulsar system J0737-3039, in which one of the pulsars is spinning rapidly with a period of 23 ms.

\end{document}